\documentclass[preprint]{aastex61}

 \usepackage{graphicx}
 \usepackage{amsmath}
 \usepackage{epstopdf,epsfig}
 \usepackage{epsf}
 \usepackage{latexsym}
 \usepackage{amssymb}
 \usepackage{longtable}

\shorttitle{Declining solar fields: Magnetospheric response}
\shortauthors{Ingale et al.}

\begin{document}

\title{The response of the terrestrial bow shock and magnetopause 
to the long term decline in solar polar fields}

\correspondingauthor{Madhusudan Ingale}
\email{mingale@prl.res.in}

\author[0000-0002-9448-1794]{Madhusudan Ingale}
\affiliation{Astronomy \& Astrophysics Division, Physical Research Laboratory, Ahmedabad 380 009, India.}   

\author{Janardhan, P.}
\affiliation{Astronomy \& Astrophysics Division, Physical Research Laboratory, Ahmedabad 380 009, India.}  

\author{Fujiki, K.}
\affiliation{Institute for Space-Earth Environmental Research, Nagoya, Japan}

\author{Susanta Kumar Bisoi}
\affil{Key Laboratory of Solar Activity, National Astronomical Observatories, \\
Chinese Academy of Sciences, Beijing 100012, China.}

\author{Diptiranjan Rout}
\affiliation{Astronomy \& Astrophysics Division, Physical Research Laboratory, Ahmedabad 380 009, India.} 



\begin{abstract}

The location of the terrestrial magnetopause (MP) and it's 
subsolar stand-off distance depends not only on the solar wind 
dynamic pressure and the interplanetary magnetic field (IMF), 
both of which play a crucial role in determining it's shape, 
but also on the nature of the processes involved in the 
interaction between the solar wind and the magnetosphere.  
The stand-off distance of the earth's MP and bow shock (BS) 
also define the extent of terrestrial magnetic fields into 
near-earth space on the sunward side and have important 
consequences for space weather.  However, asymmetries due to the 
direction of the IMF are hard to account for, making it nearly 
impossible to favour any specific model over the other in 
estimating the extent of the MP or BS. Thus, both numerical 
and empirical models have been used and compared to estimate 
the BS and MP stand-off distances as well as the MP shape, in 
the period Jan. 1975$-$Dec. 2016, covering solar cycles 
21--24. The computed MP and BS stand-off distances 
have been found to be increasing steadily over the past 
two decades, since $\sim$1995, spanning solar cycles 23 and 24.  
The increasing trend is consistent with earlier reported 
studies of a long term and steady decline in solar polar 
magnetic fields and solar wind micro-turbulence levels.  
The present study, thus, highlights the response of the 
terrestrial magnetosphere to the long term global changes in 
both solar and solar wind activity, through a detailed study of the 
extent and shape of the terrestrial MP and BS over the past 
four solar cycles, a period spanning the last four decades.  

\end{abstract}

 
\section{Introduction} \label{S-Intro}
It is well known that energetic events on the sun such as 
CMEs, flares, prominences, and high speed solar wind streams, 
gives rise to geomagnetic disturbances on the earth. Even during 
events known as solar wind disappearance events 
\citep{BaJ03, JaF05, JaF08, JDM08}, when the earth was engulfed 
by the extremely low densities observed at 1 AU ($<$ 0.1 
$\textrm{cm}^{-3}$) for periods exceeding 24 hours, the earth's 
magnetosphere and the bow shock (BS) expanded dramatically. During 
the well known and studied 11 May 1999 disappearance event, it was 
estimated that the bow shock moved outward to distances beyond 45 
earth radii (R$_{E}$) \citep{FaC01}, compared to their normal 
value of $\sim$14 R$_{E}$. recent study \citep{RoC17} has 
shown that the lateral extent of the earth's magnetopause (MP) plays 
a crucial role in determining the geoeffectiveness of solar wind outflows, 
with flows deviating by more than 6${^{\circ}}$ from the radial 
direction being non-geoeffective, as such flows will entirely miss 
the MP. Also, the BS and MP are essential in determining the 
behaviour of the magnetosheath, which plays a major role in 
the solar wind $-$ magnetosphere coupling \citep{LML11}. 
The importance of the size and shape of the MP and BS cannot therefore 
be underestimated as they play a key role in space weather studies 

It is known that the solar wind dynamic pressure and the IMF 
play a crucial role in determining the shape of the earth's 
magnetosphere, which is basically parametrized by the position 
and the shape of the BS and MP.  The BS forms in the upstream 
region of the magnetosphere, followed by the magnetosheath, 
bounded below by the MP.  The extent of the BS and MP can be 
known by estimating stand-off distances of the BS and MP.  
\citet{McA13} calculated the canonical stand-off distance 
of the BS, which is about 11 earth radii (R$_{E}$), for the 
period 2009 to 2013, covering the minimum of cycle 23 to 
the early rise phase of cycle 24, compared to about 10 R${_{E}}$ 
for the period 1974 to 1994, covering cycles 21--22. These 
changes are in keeping with the observed decline in solar 
wind dynamic pressure from $\sim$2.4 nPa (1974$-$1994) to 
$\sim$1.4 nPa (2009$-$2013).  The cycle 23 minimum in 
2008$-$2009 experienced the slowest solar wind with the 
weakest solar wind dynamic pressure and IMF as 
compared to the earlier three cycles and observations by 
{\it{Ulysses}}, the out-of-ecliptic spacecraft which explored 
the mid and high latitude heliospheric solar wind, have also 
reported a significant global decrease in the solar wind dynamic 
pressure and in the IMF during the minimum of cycle 23, as 
compared to the minima of the earlier two cycles 
\citep{RWP01, McE03, JRL11}. The changing shape of the MP and 
BS with time is therefore important in understanding space weather 
and in planetary exploration because much like the earth, other
planetary magnetosphere would have also undergone changes in 
their MP shape as a result of the observed global reduction 
in solar wind dynamic pressure. 

Equally, the long term variability in the solar magnetic fields 
can induce changes in the terrestrial magnetosphere, with the 
solar wind providing the complex link through which the effect 
is mediated. It is now well established that the near-earth space 
environment, at 1 AU, is strongly linked to the changes in the 
cyclic magnetic activity of the sun, driven essentially by the 
magnetic changes occurring in it's interior.  This link has been 
of particular interest to the solar and space science community 
due to the peculiar behavior seen in solar cycles 23 and 24 and 
in view of the long term changes taking place on the sun and in 
the solar wind \citep{JBG10, JaB11, BiJ14}.  The solar cycle 24, 
was preceded not only by one of the deepest minima in the past 100 
years but the peak smoothed sunspot number (V2.0) was $\sim$116 
in April 2014, making it the weakest sunspot cycle since cycle 14, 
which had a smoothed peak sunspot number (V2.0) of $\sim$107 in 
February 1906.  It must be clarified here that as of July 2015 a 
revised and updated list of the (Wolf) sunspot numbers has been 
adopted, referred to as V2.0 \citep{CLe16, Cli16}. Also, the 
cycle 24 is the third successive cycle in a trend of diminishing 
sunspot cycles.

Our studies \citep{JaB11, BiJ14, JaB15} have shown a steady 
and continuous decline of solar high-latitude photospheric 
fields since mid-1990's, and also in solar wind micro-tubulence 
levels in the inner heliosphere, spanning heliocentric 
distances from 0.2 to 0.8 AU \citep{BiJ14b}, in sync with photospheric 
magnetic fields. The long term declining trends seen in both 
photospheric magnetic fields and solar wind micro-turbulence 
levels over the entire inner-heliosphere, coupled with 
the unusually deep solar minimum in cycle 23 and the very 
unusual solar polar field conditions in cycle 24 \citep{GYa16}, 
implies that these changes would directly affect the size and 
shape of the terrestrial magnetosphere.  By estimating the 
variations in the stand-off distances of the 
BS and MP, one can actually quantify the effect of solar 
wind dynamic pressure and IMF on the earth's magnetosphere and 
in turn link it to the solar cycle activity.

In the present paper, we have examined the solar wind dynamic 
pressure and the IMF at 1 AU over the last four decades, from 
1975$-$2016 and estimated the stand-off distances of the BS 
and MP in order to study the behaviour of earth's magnetosphere 
over time. In addition to the direct dependence of the stand-off 
distance on solar wind dynamic pressure, it has long been 
predicted and observed that the location and shape of the BS 
and MP depends on various solar wind conditions \citep{SSA66, 
Fai71, CLy95, VeK99, FaC01}.  However, accurate theoretical as 
well as observational models for the BS and MP do not exist at 
present. We have therefore used both empirical and numerical models 
together to exploit this dependence and estimate the position of 
BS and MP as a function of time. 

Earlier studies of sunspot activity reveal periods like the Maunder 
minimum (1645-1715) when the sunspot activity was extremely low 
or almost non-existent.  Using records of ${^{14}}$C from tree rings 
over the past 1000 solar cycles or 11,000 years, \citep{USK07}
have identified 27 such prolonged or grand solar minima, each 
lasting on average $\sim$6--7 solar cycles.  The recent observations 
of anomalies in solar cycle activity in solar cycle 23 and 24, as 
mentioned earlier, have caught the attention of many researchers 
who have predicted future solar cycle activity to be heading towards 
a Maunder minimum like situation \citep{ZPo14,ZGk15,San16}.  
From a study of decadal group sunspot number, a maximum group 
sunspot number of $\sim$60 was predicted \cite{UsH14}, just prior 
to the onset of the Maunder minimum. Another study \citep{JaB15}, 
predicted a solar maximum for cycle 25, of 62$\pm$12 similar to 
conditions prior to the onset of the Maunder minimum. Recent reports 
have claimed that the sun may move into a period of very low sunspot 
activity comparable with the Dalton \citep{ZPo14} or even the Maunder 
minimum \citep{San16, ZGk15, JaB15}.  It is therefore necessary 
to access the possible impact of such unusually low solar 
activity on the earth's magnetosphere.

\section{Observations}
\subsection{Solar photospheric magnetic fields}

%
%
\begin{center}
	\protect\begin{figure}[ht]
		\vspace{0.2cm}
		\plotone{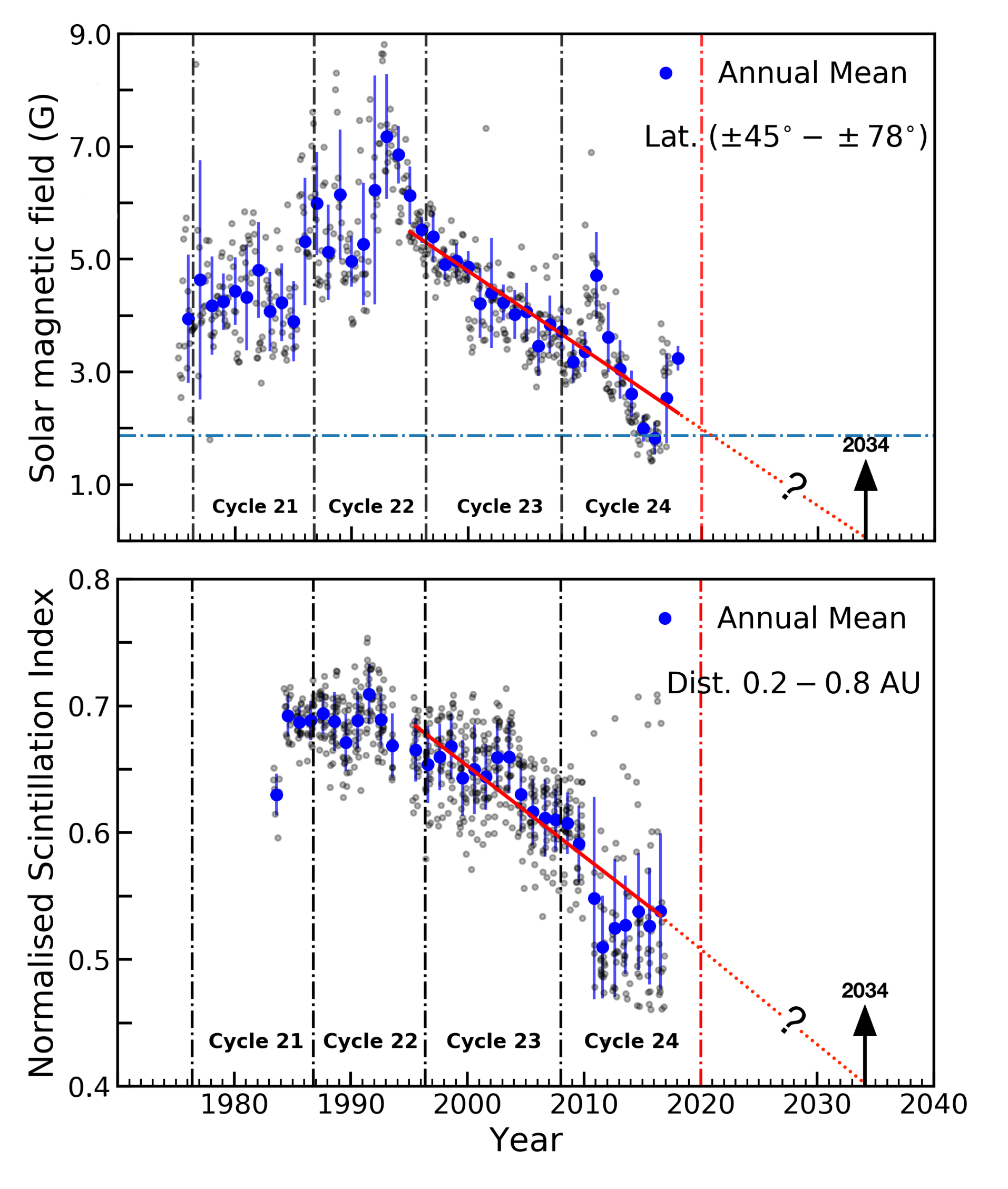}
		\caption{Photospheric magnetic fields in the latitude range $45^{\circ}-€"78^{\circ}$ for 
			the period  Feb1975$-$Dec2016 (upper
			panel) and solar wind micro-turbulence levels from 327 MHz 
			IPS observations in the period 1983$-$2016 and in the distance 
			range 0.2$-$0.8 AU (lower panel).  The filled gray dots in 
			both panels are actual measurements of magnetic fields (top) 
			and solar wind micro-turbulence (bottom), while the filled blue 
			circles are annual means shown with 1$\sigma$ error bars.  The 
			solid red line in both panels is a best fit to the declining 
			trends for the annual means while the dotted red lines are 
			extrapolations of the best fit until 2034 for the photospheric 
			fields (top) and the IPS observations (bottom).  The '?' and 
			the horizontal light blue dot-dashed line at 1.8 G are explained 
			in the text.  The vertical red dashed line in both panels 
			indicates the expected minimum of the current solar cycle 
			24 in 2020.}
		\label{fig1} 
	\end{figure}
\end{center}
%
Solar activity has been steadily decreasing over the past two 
decades and Figure \ref{fig1} shows the decline observed in both
solar photospheric magnetic fields (upper panel) and solar 
wind micro-turbulence levels (lower panel).  The upper panel 
uses observations for the period Feb.1975$-$Dec.2016 in the 
latitude range $45^{\circ}$-$78^{\circ}$.   Magnetic fields were 
computed using synoptic magnetograms from the National Solar 
Observatory, Kitt Peak (NSO/KP), the Synoptic Optical 
Long-term Investigations of the Sun (NSO/SOLIS) facility 
and the Global Oscillation Network Group (GONG).  
Each synoptic magnetogram used, is available in standard FITS 
format and represents one Carrington rotation or 27.2753 day 
averaged photospheric magnetic fields in units of Gauss. 
Further details about the computation of magnetic fields can 
be referred to in \cite{JBG10}.  

The filled grey dots in Fig.\ref{fig1} are actual measurements 
while the filled blue circles are annual means shown with 
1$\sigma$ error bars.  The solid red line is a least square 
fit to the declining trend for the annual means while the 
dotted red line is an extrapolation of the best fit until 2034, 
when the high latitude field strength will presumably drop to 
zero.  The least square fit to the magnetic field observations 
is statistically significant with a Pearson correlation coefficient 
of $r = 0.91$, at a significance level of $99\%$. It is clear 
from Fig.\ref{fig1} that the steady decline in the high 
latitude photospheric magnetic field strength has been continuing 
since $\sim$1995 and it has dropped by $\sim$40\% from its peak 
value in the period 1995$-$2016.   

An estimate of the polar or high latitude field strength at the 
minimum of a given solar cycle can be used to predict the strength 
of the next cycle maximum \citep{CLi11}.  An earlier study
\citep{JaB15}, had estimated the value of the high-latitude solar 
magnetic field in 2020, the expected minimum of the current solar 
cycle 24 to be $1.8 \pm 0.08$ G.   Using this value, shown in 
the upper panel of Fig.\ref{fig1} by a horizontal dot-dashed line, 
a sunspot maximum of 62$\pm$12 was predicted for cycle 25 
on the old un-revised sunspot count scale.

\subsection{Solar wind micro-turbulence levels}

IPS measurements essentially provide one with an 
idea of the large scale structure of the solar wind 
\citep{ACK80, ABJ95}.  Early, IPS measurements however, were 
employed in determining angular sizes of radio sources
\citep{RHe72,JAl93}. More recent observations have 
provided deep insights into the global structure of the solar 
wind and heliospheric magnetic field (HMF) all the way out to 
the solar wind termination shock at $\sim$90 AU, where 1 AU 
is the sun-earth distance \citep{FuT16}. 

The lower panel in Fig.\ref{fig1} shows the decline in 
micro-turbulence levels as measured by 327 MHz interplanetary
scintillation (IPS) observations for the period 1983 to the end 
of 2016 and in the distance range 0.2$-$0.8 AU.  These 
measurements were made using  the three station IPS observatory of the 
Institute for Space-Earth Environmental research (ISEE), Nagoya, 
Japan.  As in the upper panel, the  filled gray dots in the 
lower panel of Fig. \ref{fig1} are actual measurements of 
scintillation index for a number of compact, point-like 
extra-galactic radio sources, normalized in a manner such that 
they should show a scintillation index of unity ({\it{for 
more details see \citet{JaB11}}}).  The filled blue circles 
are annual means shown with 1$\sigma$ error bars.  The solid 
red line is a least square fit to the declining trend for the 
annual means with a Pearson correlation coefficient of 
$r = 0.93$, at a significance level of $99\%$.  The dotted 
red line is an extrapolation of the best fit until 2034. 

The implication of the scintillation index dropping to 0.4 by 
2034, if the decline continues, is that a strongly scintillating 
point-like, extra-galactic Radio source at 327 MHz will appear 
to scintillate like a much weaker and extended source having 
an angular diameter of $\sim$210 mas.  This is due to the 
significant decrease in the rms electron density fluctuations 
$\Delta$N in the solar wind over time.  As can be seen from 
the lower panel of Fig.\ref{fig1}, the scintillation level, as 
of Dec. 2016, has already dropped to around 0.5, equivalent to 
IPS observations of a source with an angular diameter of 
$\sim$150 mas. Details of the scintillation levels expected 
from 327 MHz IPS observations of radio sources having different 
angular diameters can be seen in \citet{JaB11}.  The vertical 
red dashed line in both panels of Fig. \ref{fig1} is marked 
at the expected minimum of solar cycle 24 in 2020, until which 
time it will be reasonable to assume that the decline will continue.

The decline seen in both solar photospheric magnetic fields and 
solar wind micro-turbulence levels begs the question (indicated 
by a '?' in both panels of Fig.\ref{fig1}) as to whether we are
headed towards a Maunder type "Grand" solar minimum beyond cycle 
25 wherein, the sun was devoid of sunspots in the period 1645$-$1715.  
In fact, recent theoretical modeling of sunspot number counts 
derived using the cosmogenic isotope ${^{10}}$Be, retrieved from 
deep polar ice cores that date back 140 thousand 
years, suggests the onset of a grand solar minimum in the 
period 2050$-$2200 \citep{San16}, while another study 
\citep{ZGk15} suggests significantly reduced levels of solar 
activity starting beyond cycle 25 lasting up to 2100.

\section{Data and methodology} \label{sec:dat}
%
\begin{center}
	\protect\begin{figure}[ht]
		\vspace{0.3cm}
		\plotone{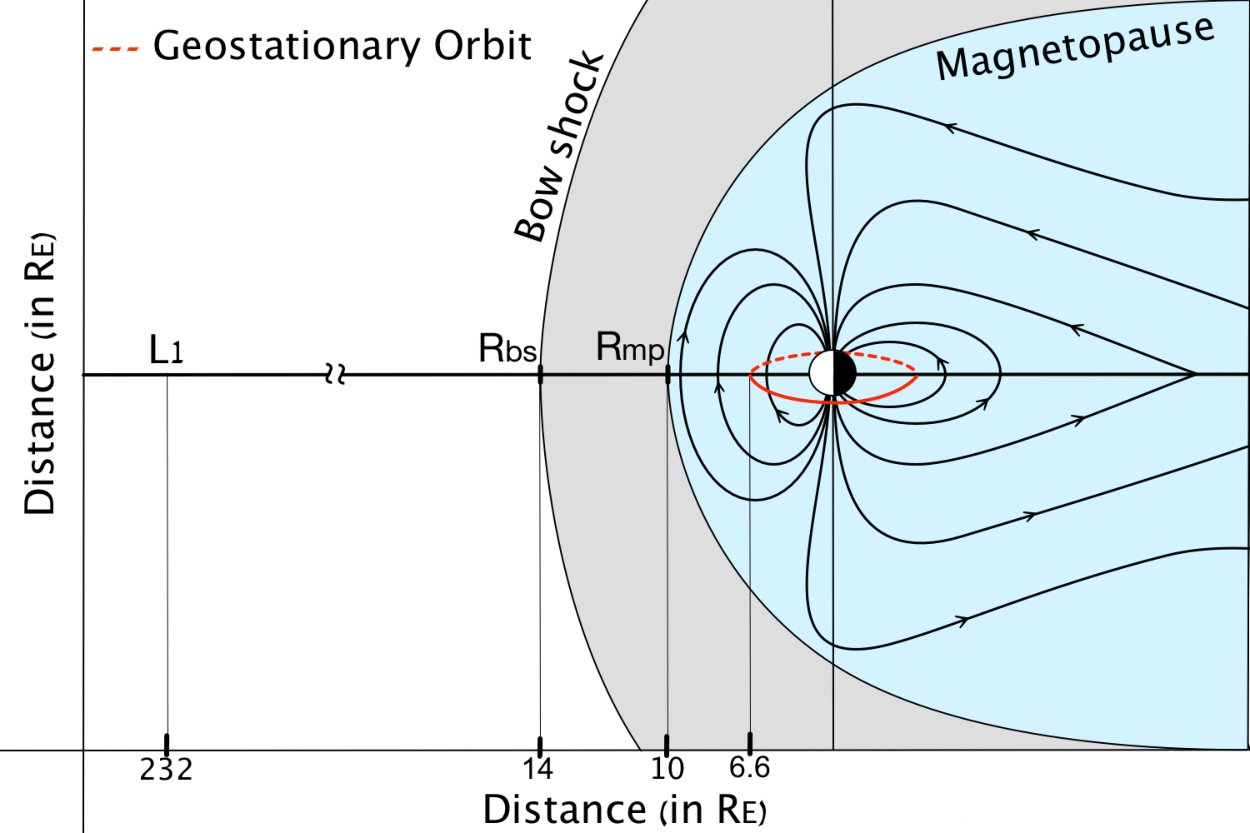} 
		\caption{A schematic of the stand-off distance of the 
			MP ($R_{\textrm{mp}}$) and BS ($R_{\textrm{bs}}$)
			in the GSM coordinate system.  The dotted red circle 
			in the equatorial plane represents the geostationary 
			orbit at 6.6 earth radii. The L1 Lagrangian point of 
			the sun-earth system is at 232 earth radii.}
		\label{fig2}
	\end{figure}
\end{center}
%
Figure \ref{fig2} shows a schematic representation (not to scale) 
of the position and the shape of the MP and BS in the GSM 
coordinate system wherein, the earth is considered 
to be at the origin. The x-axis is along the sun-earth line, 
the z-axis is perpendicular to the plane of the earth's 
orbit. $R_{\textrm{bs}}$ and $R_{\textrm{mp}}$ represent
stand-off distances of the BS and MP, respectively. The nominal 
positions of the stand-off distances of the MP and BS at 10 
and 14 earth radii (R${_{E}}$) respectively, are indicated.  
Also shown, by a red circle at 6.6 R${_{E}}$, is the 
geostationary orbit in the earth's equatorial plane and the 
location of the Lagrangian point, L1, of the sun-earth system 
at 232 R${_{E}}$.

In order to compute the stand-off distance of the BS and the MP, 
we used daily averaged data of solar wind proton density, solar wind velocity, and IMF from Jan. 1975$-$Dec. 2016. The solar wind dynamic pressure was then derived using solar wind proton density ($N_p$) 
and velocity($v_{\textrm{sw}}$).  The data sets were obtained from 
the OMNI data base, a compilation of near-earth magnetic field data 
and various other plasma parameters from several spacecraft at 
geocentric or L1 orbit which have been extensively cross compared 
and normalized (http://gsfc.nasa.gov/omniweb). In the OMNI data 
base for high resolution data (1-min and 5-min average), interpolations 
are usually performed on the phase front normal directions (for gap 
intervals of less than 3 hours), and the time shift (for gap intervals 
of less than one hour). For the purpose of this study we used low 
resolution (daily average) data, for which no interpolation was performed.
Therefore, after obtaining the OMNI data set, we replaced bad or missing 
values by the method of index aware interpolation, where the time of 
observation serves as index.

Various numerical/analytical as well as empirical models have 
been developed and used to estimate the location and shape of the 
MP and BS. Numerical models \citep{NSa91, CLy95, PeS95, EWi97, CCa03, GHu07} 
in general are evaluated using the condition of pressure balance 
between the solar wind dynamic pressure and the pressure due to 
the earth's dipole magnetic field \citep{CFe31, ZRo59, Bea60, 
SBr62, MBe64, Ols69}. While most of the empirical models,  
\citep{Fai71, For79, SLR91, ShC97, ShS98, BoE00, LiZ10, JNS12} 
with few exceptions \citep{WaS13, SGo15}, assume a functional 
form for the MP and then estemate the corresponding free parameters 
using available MP crossing database.

Each approach has several limitations, e.g., numerical/analytical 
models often use an impermeable, infinitely conducting MP as an 
obstacle, which is far from reality. On the other hand most 
of the empirical models are restricted to low latitudes. Also, 
models that use upstream parameters to describe the location and 
the shape of the MP and BS, implicitly assume proportionality 
between upstream parameters and their downstream values. This in 
turn, may lead to significant inaccuracies in cases of extreme 
solar wind conditions.

While empirical models provide the average values of MP and BS, 
which are in good agreement with their observational values, 
it is important to note that since our principle aim is to 
study the response of the MP and BS to the long term and steady 
decline in activity seen on the sun and in th solar wind 
\citep{JBG10, JaB11, BiJ14,JaB15}, our approach is 
concentrated on estimating the long term trend and changes 
in the MP and BS stand-off distance and the MP shape. 
Therefore, in what follows we estimate the MP stand-off distance 
using models due to \citet{LiZ10}, and \citet{LuL11}, as being
representative of empirical and numerical models, respectively.  
Similarly, the BS stand-off distance has been estimated using 
\citet{JNS12} and \citet{CCa03} as being representative of empirical 
and numerical models, respectively.

\subsection{MP stand-off distance}
The boundary between the solar wind and magnetosphere can be derived 
using the pressure balance condition. This condition supposes that the MP 
and BS can be described as second order surfaces i.e. a surface described
by an algebraic equation of degree two.  Most of the time 
therefore, elliptic or parabolic functional forms are used to represent 
the shape of the MP and BS. However elliptic or parabolic functional forms 
are generally, not appropriate in describing the MP tail and \citet{ShC97} 
proposed the following functional form,

\begin{equation}
	r = r_{mp} \left(\frac{2}{1 + \cos\theta}\right)^{\alpha} \,\,\, R_E. 
	\label{mp_position}
\end{equation}

Equation (\ref{mp_position}) has two parameters ${\it{viz.}}$ 
r${_{mp}}$, the MP stand-off distance and $\alpha$, the flaring 
parameter. Angle $\theta$ is the solar zenith angle (the angle 
between the Sun-Earth line and the radial direction) of the point 
of interest \citep{SSo02}. Eq. (\ref{mp_position}) represents an
open MP tail for $\alpha > 0.5$ and a closed MP tail for 
$\alpha < 0.5$.

We now briefly describe the models we used in the 
calculations of the of the MP stand-off distance and MP shape.
The first one is due to \citet{LiZ10} which represents an  
empirical approach, whereas second model, \citet{LuL11} is 
representative of a numerical approach.

\subsubsection{Empirical model (Using Lin et al., 2010)}

\citet{LiZ10}, abbreviated as L10 hereafter, extended the 
assumptions of \citet{ShS98} to address asymmetries and 
indentations near the polar cusps.  Employing a database 
of nearly 2708 MP crossings from observations by Cluster, 
Geotail, Goes, IMP8, Interball, LANL, Polar, TC1, THEMIS, 
WIND and Hawkeye, along with corresponding solar wind 
parameters from ACE, Wind and OMNI, they obtained a model 
for the MP which was parametrised by the solar wind dynamic 
and magnetic pressure (P${_{d}}$ + P${_{m}}$), IMF B${_{z}}$ 
and dipole tilt angle ($\psi$), which is the dipole magnetic 
latitude of the subsolar point. Based on the relation between 
P${_{d}}$ and r${_{mp}}$, the influence of IMF B${_{z}}$ on 
r${_{0}}$ \citep{ShS98} and the saturation effect of a 
southward IMF B${_{z}}$ on r${_{mp}}$ \citep{YaC03}, L10 
expressed the stand-off distance for MP as:

\begin{equation} 
r_{mp} = a_0(P_d + P_m)^{a_1}\left(1 + a_2 \frac{\textrm{exp}(a_3 B_z)-1}
{\textrm{exp}(a_4 B_z) -1}\right) \,\,\, R_E
\label{L10} 
\end{equation}

The coefficients ($a_0$, $a_1$, $a_2$, $a_3$ and $a_4$) in 
equation (\ref{L10}) are obtained by using nonlinear multi 
parameter fitting (Levenberg $-$ Marquardt method) based on 
observations of 247 MP crossings that were found near the 
stand-off distance. The coefficients are listed in Table$-$2 
of L10. 

To obtain the MP shape, L10 expanded the eq. (\ref{mp_position}) as,

\begin{equation}
r = r_{mp} \left \{\mathrm{cos} \frac{\theta}{2} + a_5 \cdot \mathrm{sin}
(2 \theta) [1 + \mathrm{exp}(-\theta)] \right \}^{\beta} \,\, R_E.
\label{mp_shape_L10}
\end{equation}
Where the factor $1-exp(-\theta)$ smooths out the MP shape near 
the subsolar point.  The asymmetries and indentations are introduced 
through the azimuthal angle $\phi$, the angle between the projection 
of r in Y-Z plane and the direction of positive Y axis.  The flaring 
parameter $\beta$ (given by equation (5) of L10) is,
\begin{equation}
\beta = \beta_0 + \beta_1 \mathrm{cos}(\phi) + \beta_2 \mathrm{sin}(\phi)
+ \beta_3 \mathrm{sin}^2(\phi).
\label{beta}
\end{equation}

We considered the simpler case: $\phi=0$ (meridional plane) 
for which equation (\ref{beta}) reduces to $\beta = \beta_0 + 
\beta_1$. $\beta_0$ and $\beta_1$ (eq. 16 and 17 of L10) are 
obtained using observations of 422 MP crossings. The relevant 
values of the parameters are listed in table$-$6 of L10.

The L10 model (eqs \ref{L10} and \ref{mp_shape_L10}) yields good 
results in predicting MP stand-off distance and MP shape 
respectively. When compared with the observed low latitude 
MP crossings, the standard deviation of the L10 model is the 
least (0.54 $R_E$) among several other models.
 
\subsubsection{Numerical model (using Lu et al., 2011)}
The second model we used is due to \citet{LuL11}, abbreviated as 
L11 hereafter, based on global MHD simulations to estimate the MP 
stand-off distance and the MP shape. L11 analyses the relation 
between the MP and the IMF B${_{z}}$ and P${_{d}}$ using numerical results 
from a global MHD model Space Weather Modelling and Framework 
(SWMF), a framework for physics-based space weather simulations 
\citep{ToS05}. A streamline technique was used to identify the location 
and shape of the MP. The functional form of \citet{ShC97} was extended 
to describe the global MP size and shape using the method of multi-parameter
fitting. L11 included azimuthal asymmetry via ($\phi$) and extended the 
functional form in eq. (\ref{mp_position}). The dayside MP is given by, 

\begin{equation}
r = r_{mp} \left(\frac{2}{1 + \mathrm{cos}\theta} \right)^{\alpha + \beta_1 
\mathrm{cos}\phi} \,\, R_E.
\label{mp_shape_L11}
\end{equation}

Where $\beta_1$ characterises the azimuthal asymmetry with respect to 
$\phi$. Using fitting results from \citet{ShC97}, the relationship 
between the (r${_{0}}$, $\alpha$, $\beta_1$) and solar wind conditions 
(P${_{d}}$, B${_{z}}$) was evaluated. The multiple parameter fitting 
results in the following best-fit functions (eq., 18, 19, 20 of L11):

\begin{equation}
    r_{mp} = 
\begin{cases}
    (11.494 + 0.0371 B_z) P_d^{-1/5.2}, \,\, B_z \geq 0 \\
    (11.494 + 0.0983 B_z) P_d^{-1/5.2}, \,\, B_z < 0
\end{cases}
\label{r0_L11}
\end{equation}

\begin{equation}
    \alpha = 
\begin{cases}
    (0.543 - 0.0225 B_z + 0.00528 P_d + \\ 0.00261 B_z P_d), \,\, B_z \geq 0 \\
    (0.543 - 0.0079 B_z + 0.00528 P_d + \\ 0.00019 B_z P_d), \,\, B_z < 0
\end{cases}
\label{alpha_L11}
\end{equation}

\begin{equation}
    \beta_1 = 
\begin{cases}
    (-0.263 + 0.0045 B_z - 0.00924 P_d - \\ 0.00059 B_z P_d), \,\, B_z \geq 0 \\
    (-0.263 - 0.0259 B_z - 0.00924 P_d + \\ 0.00256 B_z P_d), \,\, B_z < 0
\end{cases}
\label{beta_L11}
\end{equation}

The model due to L11 yields good matching when compared with 
the high and low latitude MP crossings.

\subsection{BS stand-off distance} \label{subsec:BS}
The shape and the location of the BS mainly depends on the 
location of the MP as well as various solar wind parameters, 
for example the magnetosonic and Alfv\'en Mach numbers. In 
general, the shape of the BS is assumed to be a paraboloid 
along the Earth $-$ Sun line. We now briefly describe the 
models we used in the calculations of the BS stand-off distance.
The first model, \citet{JNS12} represents an empirical approach 
and the second one, \citet{CCa03}, is a representative of 
numerical approach.
 
\subsubsection{Empirical model (Je\'linek et al. 2012)}
\citet{JNS12}, hereafter abbreviated as J12, investigated the solar 
wind (SW), magnetosheath (MSH) and magnetosphere (MS) using measurements 
from the THEMIS spacecraft between March 2007 and September 2009. The 
orbits of the five THEMIS spacecraft spans all three regions of interest: 
SW, MSH and MS. The ACE spacecraft at L1 was used as a solar wind monitor. 
The ratio of the measurements of the magnetic field and density from THEMIS 
with those from ACE enables one to identify SW, MSH and MS and the 
boundaries between these regions for the entire day-side part.  Since BS and 
MP are often described as second order surfaces, J12 expected a parabolic 
shape for both of the boundaries that responds to the upstream dynamic 
pressure, P${_{d}}$ as,
\begin{equation}
r \sim P_d^{-1/\epsilon}.
\end{equation}

J12 assumes rotationally symmetric BS and MP around the $X_{GSM}$. 
An analytic expression for the BS and MP is used and the free 
parameters are determined using least square fitting to the full 
data set. However the MP model due to \citet{JNS12} does not 
include the effect of the dipole tilt angle, which is the dipole 
magnetic latitude of the subsolar point, and is therefore severely 
restricted to the low latitudes, whereas the L10 model described in 
\S{3.1.1} is applicable in more general situations. We therefore 
only considered the model of the BS stand-off distance from the 
\citet{JNS12}, given by.

\begin{equation}
r_{bs} = 15.02 P_d^{-1/6.55} \,\,\, R_E.
\label{JRbs}
\end{equation}

Although the equation (\ref{JRbs}) does not take into account the 
Mach number, this simple model was found in good agreement when compared 
with more than 6000 BS crossings.

\subsubsection{Numerical model (Chapman \& Cairns 2003)}

The model due to \citet{CCa03}, hereafter abbreviated as CC03, 
uses 3D ideal MHD simulations \citep{CLy95}, which in turn 
uses an impermeable infinitely conducting magnetopause 
(given by \citet{FPR91}) as an obstacle. This model is 
parametrised by P${_{d}}$, the Alfv\'en Mach number (M${_{A}}$) and 
the orientation of the IMF ($\theta_{IMF}$) with respect to 
the direction of the solar wind velocity ($v_{sw}$). \citet{CCa03} 
considers two special cases of $\theta_{IMF} = 45^{\circ}$ 
and $90^{\circ}$, of which we use $\theta_{IMF} = 90^{\circ}$ to 
estimate the BS stand-off distance given by
\begin{equation} 
r_{bs} = \left(\alpha_0 + \frac{\alpha_1}{M_A} \right)
\left(\frac{P_d}{1.87} \right)^{-1/6} \,\, \textrm{$R_E$}. 
\label{CC04}
\end{equation}
Here, ($\alpha_0, \alpha_1$) are the fitting parameters obtained by 
least square fitting to the simulated shock locations. The CC03 model 
has been compared with available spacecraft data close to the nose 
region of the BS and it was found that for the near-Earth regime 
($-20 R_E < x < 35 R_E$), the model does well in predicting the BS 
location.

\section{Results} \label{sec:results}

One of the key parameters in determining the shape and location of 
the BS and MP is their stand-off distance, which depends principally 
on P${_{d}}$ and the strength of the IMF.  Therefore, the variations 
in the stand-off distances as a function of P${_{d}}$ and IMF gives 
one a good handle in understanding the response of the earth's
magnetosphere to the changes in solar wind conditions and in turn 
to the global variability in solar activity. We now present 
and compare the results for the BS stand-off distance obtained by 
using empirical \citep{JNS12} and numerical (MHD) model \citep{CCa03},
described in (\S{3.2.1}) and (\S{3.2.2} respectively.) 

We have computed the BS stand-off distance and normalised it to its average 
value subjected to the typical solar wind conditions at 1 AU. This has 
been done to show the excursions of the BS beyond the average 
stand-off distance for cycles $21-24$. For typical solar wind
conditions at 1 AU (P${_{d}}$ $\sim$1.87 nPa, B = 7 nT, N${_{p}}$ = 6.6 \textrm{cm}${^{-3}}$; M${_{a}}$ = 8), the average BS stand-off distance according to J12 is $13 R_{E}$ and that of due to CC03 is $18 R_E$. The 
difference in the magnitude of the BS stand-off distance estimated 
by these two models can be understood in the following way. Both the models, 
%
\begin{center}
	\protect\begin{figure*}[ht!]
		\vspace{0.3cm}
		\plottwo{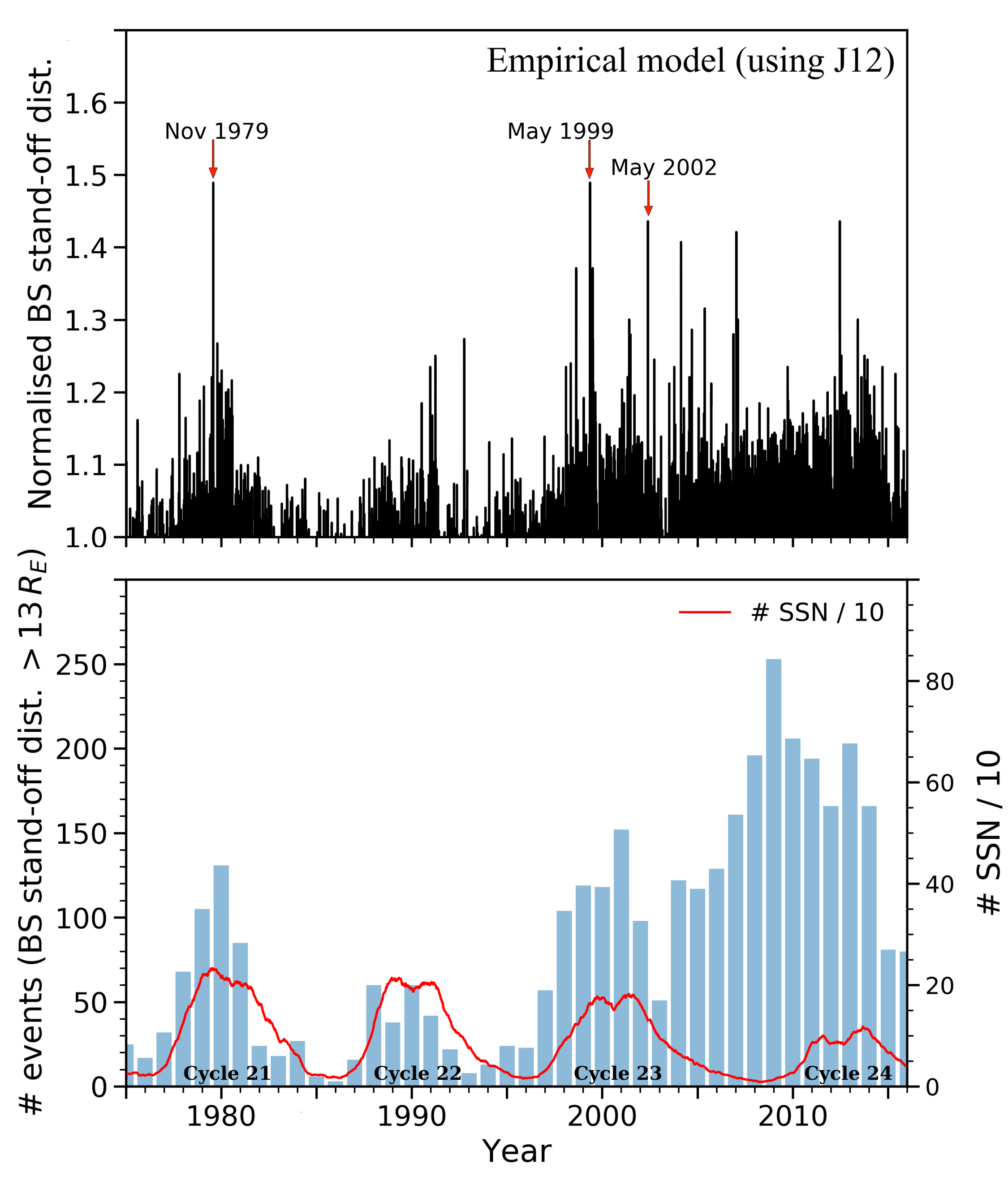}{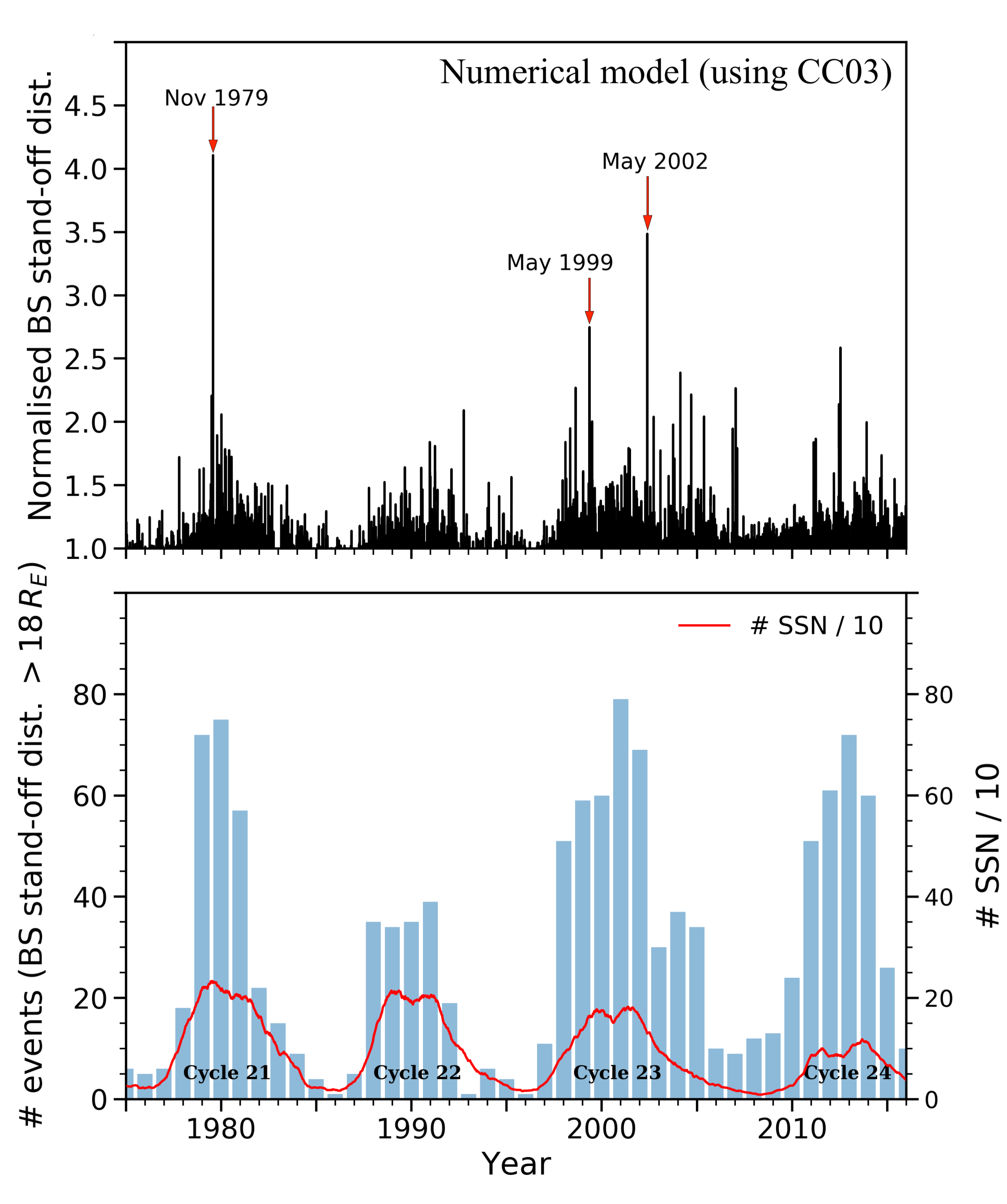} 
		\caption{Daily average of the normalized BS stand-off 
			distance between January 1975 and December 2016. Left 
			(top and bottom) panels uses J12, while Right (top 
			and bottom) panels uses CC03.  Three extreme events, 
			designated as solar wind disappearance events in the 
			literature, have been labeled with the event dates 
			(upper panels). The distribution of the number of 
			events or instances for which the BS stand-off 
			distance $>$ average stand-off distance between 1975 
			and 2016 (bottom panels).  A 12 month moving average 
			of the sunspot numbers, scaled down by a factor of 10 
			is shown over plotted  (in red) on the histogram.}
		\label{fig4} 
	\end{figure*}
\end{center}
%
%
CC03 and J12 relates BS stand-off distance with solar wind dynamic 
pressure as a power law, but with different power law indices. Also 
CC03 model includes the effect of the Alv\'en Mach number (M${_{A}}$), 
which is neglected in J12.

In Figure \ref{fig4} left panels (top and bottom) shows the results 
obtained by using an empirical model (J12, \S{3.2.1}) and right panels
(top and bottom) shows results obtained by using a numerical model 
(CC03, \S{3.2.2}). The upper panel (top left and right) of Fig. \ref{fig4} 
shows the daily average of the normalized BS stand-off distance 
from 1975 to December 2016, derived using J12 (top left) and 
CC03 (top right).  It is clear that the normalized BS stand-off 
distance, on an average, follows the eleven year solar cycle. 
Surprisingly though, such excursions of the BS well beyond 
average stand-off distance are not rare and are much more 
frequent than expected. From the upper panel of Fig.\ref{fig4} it is 
clear that, irrespective of the model used
there are a large number of cases of the BS stand-off distance 
extending well beyond the average value. Three such events are 
indicated in the Fig.\ref{fig4} (upper panel). These events have 
been well studied and are referred to as solar wind disappearance 
events \citep{BaJ03, JaF05, JaF08, JDM08} due to the extremely 
low densities observed at 1 AU ($<$ 0.1 $\textrm{cm}^{-3}$) for 
periods exceeding 24 hours.  During all three events, a sharp 
decrease in $P_d (< 0.02 ~ \textrm{nPa}$) was seen indicating 
sensitive response of the BS stand-off distance to solar wind 
conditions. 

To quantify the effect of the solar wind conditions on the BS, 
we selected events for which the BS stand-off distance 
increases more than 1$\sigma$ of the average value of the BS 
stand-off distance. The lower panel of Fig. \ref{fig4} (lower 
left and right) shows the histogram of the distribution of the 
number of events for the years between 1975 and 2016 obtained 
using J12 (lower left) and CC03 (lower right). As stated earlier, 
it is clear that BS excursions beyond average distance are not 
rare but are observed consistently in each solar cycle. However, 
there is a significant increase in the number of events since 
$\sim$1995 when solar photospheric magnetic fields began declining. 
The increase was found to be more than 40\% when compared with the 
number of events before 1995. Note that the increase in the number 
of events is irrespective of the models used.

We now turn to the discussion of the MP stand-off distance 
and the MP shape obtained by empirical \citep{LiZ10} and 
numerical (MHD) models \citep{LuL11}, described in (\S{3.1.1}) 
and (\S{3.1.2}) respectively.  The MP stand-off distance for 
the years Jan. 1975$-$Dec. 2016 is shown in Figure \ref{fig5}. 
The top panel shows the result obtained by using L10 whereas, the 
results in the lower panel are derived using L11. The grey dots are 
monthly averages of the MP stand-off distance. The blue circles 
represent annual averages shown with 1$\sigma$ error bars. The monthly 
averaged sunspot number, scaled down by a factor of 10, is shown by 
a grey solid line with the smoothed value (one year moving average) 
over plotted in blue.  It is clear from the figure that the MP 
stand-off distance is sensitive to and is modulated, with a periodicity 
of 11 years, by the solar cycle.

To remove this periodicity and investigate the trend, daily average 
of the MP stand-off distance was smoothed using a eleven year running 
mean, shown by the over plotted solid red curve in Fig. \ref{fig5}, 
which shows a clear increasing trend in the MP stand-off distance 
starting in $\sim$1995, and monotonically increasing till 2015. The 
increase is $\sim15\%$, irrespective of the models used.

We also determined the shape of the MP. Following the standard GSM 
coordinate system we computed position of the MP, $X_s = r sin(\theta)$ 
and $R_s = r cos(\theta) = \sqrt(Y^2 + Z^2)$, where $\theta$ is the 
solar zenith angle. $R_s$ is interpreted as the transverse radius. 
We refer to the plot of $R_s$ v/s $X_s$, averaged over 11 years, 
as the shape of the MP or MP shape.
%
\begin{center}
	\protect\begin{figure}[ht!]
		\vspace{0.0cm}
		\plotone{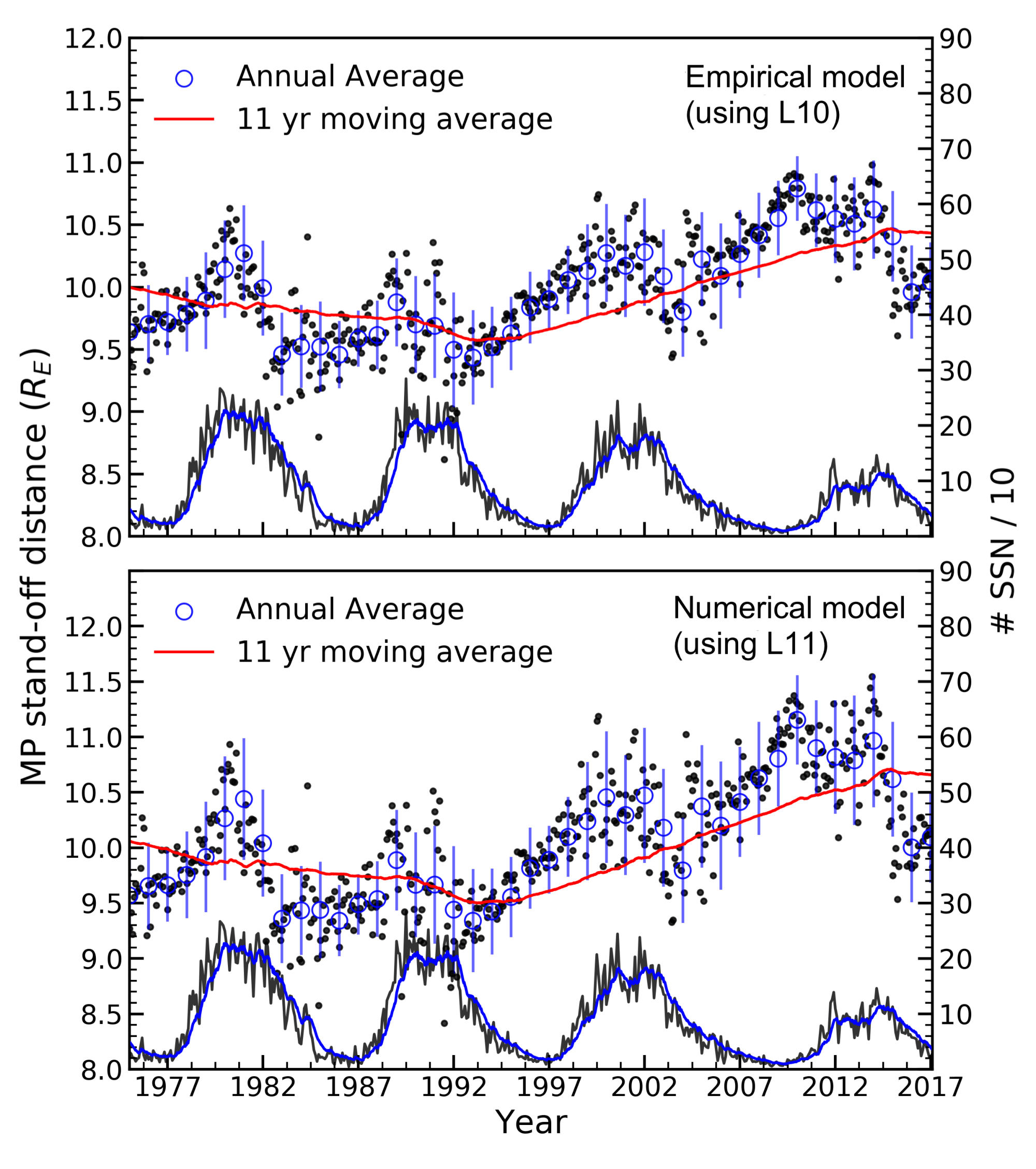} 
		\caption{Monthly averages of the MP stand-off distance 
			for the period Jan. 1975$-$Dec. 2016  (grey filled 
			circles) derived using \citet{LiZ10} (L10) (upper panel) 
			and \citet{LuL11} (L11) (lower panel). 
			The blue circles represent annual averages shown with 
			1$\sigma$ error bars. The red line is a eleven year 
			moving average of the daily average of the MP stand-off 
			distance. The monthly averaged sunspot number, scaled 
			down by a factor of 10, is shown by the solid curve 
			in grey with the smoothed value (one year moving 
			average) over plotted in blue.}
		\label{fig5} 
	\end{figure}
\end{center}
%
Figure \ref{fig6} shows the plot of $R_s$, the transverse radius 
against $X_s$ the stand-off distance along the earth-sun line for 
the day-side MP and for a solar zenith angle between $\theta$ = 0${^{\circ}}$ and $\theta$ = 90${^{\circ}}$. Upper panel of the Fig. \ref{fig6} shows the MP shape obtained using eq. (\ref{mp_shape_L10}) 
(\S{3.1.1}), whereas lower panel shows the MP shape derived using 
eq. (\ref{mp_shape_L11}) (\S{3.1.2}).
%
\begin{figure}[ht!]
	\vspace{0.1cm}
	\plotone{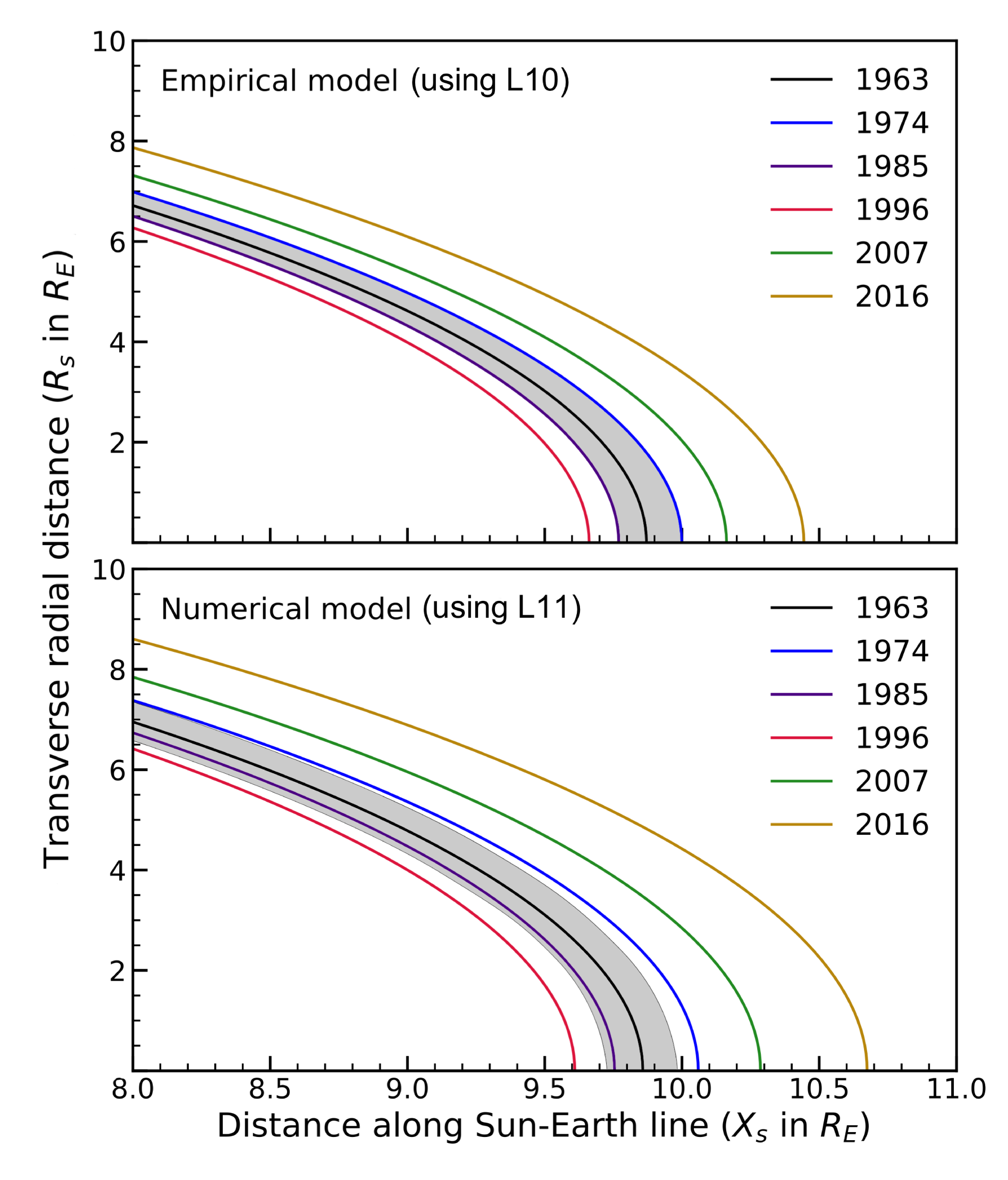} 
	\caption{The 11 year averaged MP shape shown by a 
		plot of the transverse radial distance of the MP, $R_s$ 
		against the extent of the MP along the sun-earth 
		line, $X_s$. The upper panel uses \citet{LiZ10} (L10) 
		(\S{3.1.1}) and the lower panel uses \citet{LuL11} (L11) 
		(\S{3.1.2}). The black line is the eleven year average 
		starting from 1963, with shaded gray band having a 
		1$\sigma$ width.  The blue, indigo, red, green and gold 
		curves represent 11 year averaged MP shapes for years 
		1974, 1985, 1996, 2007 and 2016, respectively.}
	\label{fig6} 
\end{figure}
%

The average MP shape is labeled by the starting year, {\it{e.g.}} 
the MP shape of 1963, shown by the black curve, refers to the MP 
shape that is averaged over the eleven years starting from 1963. 
The gray band in Fig. \ref{fig6} signifies the region of 1$\sigma$ 
around the MP shape of 1963.  It is important to note that, in case 
of L10 (upper panel), the MP shape of 1974 and 1985 falls within this 
gray band. Whereas, in case of L11 (lower panel) the average MP shape 
for the year 1974 is slightly outside the 1$\sigma$ region of the MP 
shape of 1963. However, in both cases (L10 and L11) the average MP 
shape in 1996 falls well below the average MP shape of 1974, and then 
bounces forward significantly in 2007 and continues to expand 
till 2016. It can be seen that the expansion of the MP is different 
at different MP positions but there is an overall expansion in MP 
with the maximum expansion being at the stand-off point. For the
period between 1996 and 2016,  the stand-off point expanded by nearly 
1 R${_{E}}$ from $\sim$9.6 R${_{E}}$ to $\sim$10.5 R${_{E}}$,  
irrespective of the model used. 

\section{Discussion and conclusions} 

We have carried out an extensive study based on IMF and solar 
wind data between January 1975 and December 2016. Owing 
to the complex nature of the solar wind - magnetosphere interaction, 
determining the boundaries (BS and MP) is an open question. 
At the present time accurate theoretical and observational models 
for the shape and location of the BS and MP do not exist.
However, our aim was to investigate the response of the stand-off 
distance of the BS and MP to the observed decline in solar activity - 
characterised by the change in the high latitude magnetic field 
\citep{JaB15} and micro-turbulence levels over the heliocentric 
distance between 0.2 to 0.8 AU \citep{JaB11}. 
We therefore carried out the calculations and compared the 
results for several available models. We presented the results 
for the MP stand-off distance and the MP shape using L10 
(empirical model, \S{3.1.1}) and compared them with the results 
obtained from L11 (global MHD simulations, \S{3.1.2}). For the 
BS stand-off distance we presented the results using CC03 (Global 
MHD simulations, \S{3.2.1}) and compared them with the results 
obtained by J12 (empirical model, \S{3.2.2}). We found that the 
long term trend in the BS and MP as a response to solar wind 
conditions and IMF is, in general, independent of the model used.

The stand-off distance of the MP and BS are sensitive to 
variations in P${_{d}}$  and IMF B${_{z}}$. They are also 
affected by the Alfv\'en and magnetosonic Mach numbers. The 
angle between the magnetic field and the solar wind velocity
vector is critical in determining the shock location of 
the BS \citep{CLy95}. The shape of the magnetopause is asymmetric 
due to the cusp in the polar regions, this asymmetry can be 
accounted if the dipole tilt angle ($\psi$) is taken into 
consideration \citep{LiZ10, SGo15}. To simplify the analysis 
we considered MP and BS symmetric about the sun-earth line 
(i.e. the X-axis of the GSM coordinate system) and neglected the 
dipole tilt angle. For calculating BS stand-off distance using 
CC03 we kept IMF angle ($\theta$) fixed at $90^{\circ}$.

A decrease in P${_{d}}$ and IMF causes an expansion 
of the BS and MP, resulting in an increase in their sub-solar 
distances. The stand-off distance of the MP, in general, 
exhibits a power law dependence on the dynamic pressure, with 
power law index $\sim-1/6$ \citep{MBe64, PRu96}. A self-similar 
scaling suggests an identical power law dependence for the 
stand-off distance of the BS \citep{CLy96}. However, the power 
law index found in several empirical as well as numerical studies 
is little less than $-1/6$. This reduced value might be the effect 
of the pressure due to earth's dipole field \citep{JNS12}. It is 
worth noting here that both approaches numerical/analytical and 
empirical either implicitly or directly include the earth's dipole 
field. \citet{ZhW14} have shown that the earth's dipole moment has 
been decaying over the past 1.5 centuries. Assuming linear rate of 
decay to persist their results suggests, the average stand-off 
distance of the MP would move $\sim 0.3 R_E$ towards the earth per 
century. We are looking for trend in the MP stand-off distance 
averaged over the eleven years and computing eleven year average 
of the MP shape so the modification to the MP shape at subsolar 
point due to the change in the pressure caused by declining dipole 
field will be very small and hence negligible.

The main conclusions of the paper are:
\begin{enumerate}
\item Corresponding to the observed decrease of $\sim$40\% in P${_{d}}$, 
a steady increase of $\sim$15\% was observed in the stand-off 
distance of the MP which can be attributed to the power law 
dependence.

\item We also observed a significant increase of more than 40\% in the 
number of events (after 1995) where the stand-off distance of 
the BS exceeded the average stand-off distance over the past $\sim$20 
years, when compared with the number of events prior to 1995. 
This indicates that the earth's magnetosphere is very sensitive 
to the changes in solar wind conditions.  

\item From our study of the variation in the 11 year average 
shape of MP, we found that the MP, since 1996, has undergone 
a significant expansion which is highest at the stand-off 
point and narrows down towards the transverse radius. Our result 
for the shape of the MP is consistent with the increase in the
stand-off distance of the MP reported by \cite{McA13} wherein, 
they found an increase in the stand-off distance of the MP from 
10 $R_E$ in 1974$-$1994 to 11 $R_E$  in 2007--2013.  The values 
of P${_{d}}$ and the stand-off distance of MP, from 1974$-$1994, 
were respectively, $\sim$2.9 nPa and 9.7 R${_{E}}$.  In contrast, 
these values, from 1995$-$2017, were found to be respectively, 
$\sim$2.0 nPa and 10.7 R${_{E}}$.  So our results of the increase 
in the MP stand-off distance ($\sim$9.3\%) are consistent with the 
increase in stand-off distance of MP ($\sim$10\%) reported by 
{McA13}.  Our results also showed an increase in MP shape 
($>$ 10 $R_E$) for the period from 2007$-$2016.

\item During solar minimum, photospheric high latitude magnetic 
fields extend to low latitudes and are then pulled into the 
heliosphere by the solar wind thereby, forming the 
IMF \citep{SPe93}.  The changes in the solar wind conditions 
such as decline in P${_{d}}$ and IMF strength can thus be 
interpreted as being induced by global changes in the solar 
magnetic fields. Our study underlines the causal relation 
between solar activity changes and the corresponding global 
response of the earth's magnetosphere via the variations 
quantified by the stand-off distances of the BS and MP.

\item The present work quantifies the response of the earth's 
magnetosphere via the variations in the stand-off distance of 
the MP and BS subjected to the lomg term changes in P${_{d}}$ 
and IMF.  We have found that both the P${_{d}}$ and the IMF 
have been steadily declining since $\sim$1995 with the 
reduction in their average values over the last 20 years
being $\sim$40\%.  This is consistent with the ongoing 
declining trend in high-latitude photospheric magnetic fields 
and solar wind micro-turbulence levels, both of which showed 
a decrease in their strength beginning around $\sim$1995.  

\item We find that the steady decline in high-latitude photospheric 
fields and solar wind micro-turbulence levels are still continuing 
implying low sunspot activity in future, a condition akin to the 
Maunder minimum. Using a global thermodynamic model, \cite{RiL15} 
reported the likely state of the solar corona, during the later 
period of the Maunder minimum, devoid of any large scale structure 
and driven by a reduced photospheric magnetic field strength.  
The photospheric field strength during the last two solar cycles 
has been steadily decreasing (since $\sim$1995) and the trends indicate 
that it is likely to decline in the same manner in future solar 
cycles.  This implies a state of corona with no large scale 
structure much alike Maunder minimum period, which in turn, leads 
to a highly bulged terrestrial magnetosphere with an increased 
stand-off distance for the bow shock and the magnetopause.

\end{enumerate}

Continued investigation to understand and forecast the 
influence of solar activity on the near earth environment 
and the ecosystem is therefore of considerable importance. 

\acknowledgments

This work has made use of NASA's OMNIWEB services Data System. 
The authors thank the free data use policy of the National Solar
Observatory (NSO/KP, NSO/SOLIS and NSO/GONG). JP and DR acknowledge 
the ISEE International Collaborative Research Program for support 
during this work.

\bibliographystyle{apj}

\begin{thebibliography}{64}
	\expandafter\ifx\csname natexlab\endcsname\relax\def\natexlab#1{#1}\fi
	
	\bibitem[{{Ananthakrishnan} {et~al.}(1995){Ananthakrishnan}, {Balasubramanian},
		\& {Janardhan}}]{ABJ95}
	{Ananthakrishnan}, S., {Balasubramanian}, V., \& {Janardhan}, P. 1995, \ssr,
	72, 229
	
	\bibitem[{{Ananthakrishnan} {et~al.}(1980){Ananthakrishnan}, {Coles}, \&
		{Kaufman}}]{ACK80}
	{Ananthakrishnan}, S., {Coles}, W.~A., \& {Kaufman}, J.~J. 1980, \jgr, 85, 6025
	
	\bibitem[{{Balasubramanian} {et~al.}(2003){Balasubramanian}, {Janardhan},
		{Srinivasan}, \& {Ananthakrishnan}}]{BaJ03}
	{Balasubramanian}, V., {Janardhan}, P., {Srinivasan}, S., \& {Ananthakrishnan},
	S. 2003, \jgr, 108, 1121
	
	\bibitem[{{Beard}(1960)}]{Bea60}
	{Beard}, D.~B. 1960, \jgr, 65, 3559
	
	\bibitem[{{Bisoi} {et~al.}(2014{\natexlab{a}}){Bisoi}, {Janardhan},
		{Chakrabarty}, {Ananthakrishnan}, \& {Divekar}}]{BiJ14}
	{Bisoi}, S.~K., {Janardhan}, P., {Chakrabarty}, D., {Ananthakrishnan}, S., \&
	{Divekar}, A. 2014{\natexlab{a}}, \solphys, 289, 41
	
	\bibitem[{{Bisoi} {et~al.}(2014{\natexlab{b}}){Bisoi}, {Janardhan}, {Ingale},
		{Subramanian}, {Ananthakrishnan}, {Tokumaru}, \& {Fujiki}}]{BiJ14b}
	{Bisoi}, S.~K., {Janardhan}, P., {Ingale}, M., {et~al.} 2014{\natexlab{b}},
	\apj, 793, 8
	
	\bibitem[{{Boardsen} {et~al.}(2000){Boardsen}, {Eastman}, {Sotirelis}, \&
		Green}]{BoE00}
	{Boardsen}, S.~A., {Eastman}, T.~E., {Sotirelis}, T., \& Green, J.~L. 2000,
	\jgr, 105, 23,193?23,219
	
	\bibitem[{{Cairns} \& {Lyon}(1995)}]{CLy95}
	{Cairns}, I.~H., \& {Lyon}, J.~G. 1995, \jgr, 100, 17173
	
	\bibitem[{{Cairns} \& {Lyon}(1996)}]{CLy96}
	---. 1996, \grl, 23, 2883
	
	\bibitem[{{Chapman} \& {Cairns}(2003)}]{CCa03}
	{Chapman}, J.~F., \& {Cairns}, I.~H. 2003, Journal of Geophysical Research
	(Space Physics), 108, 1174
	
	\bibitem[{{Chapman} \& {Ferraro}(1931)}]{CFe31}
	{Chapman}, S., \& {Ferraro}, V. C.~A. 1931, Terr. Magn. Atmos. Electr., 36, 77
	
	\bibitem[{{Clette} \& {Lef{\`e}vre}(2016)}]{CLe16}
	{Clette}, F., \& {Lef{\`e}vre}, L. 2016, \solphys, 291, 2629
	
	\bibitem[{{Cliver}(2016)}]{Cli16}
	{Cliver}, E.~W. 2016, \solphys, 291, 2891
	
	\bibitem[{{Cliver} \& {Ling}(2011)}]{CLi11}
	{Cliver}, E.~W., \& {Ling}, A.~G. 2011, \solphys, 274, 285
	
	\bibitem[{{Elsen} \& {Winglee}(1997)}]{EWi97}
	{Elsen}, R.~K., \& {Winglee}, R.~M. 1997, \jgr, 102, 4799
	
	\bibitem[{{Fairfield}(1971)}]{Fai71}
	{Fairfield}, D.~H. 1971, \jgr, 76, 6700
	
	\bibitem[{{Fairfield} {et~al.}(2001){Fairfield}, {Cairns}, {Desch}, {Szabo},
		{Lazarus}, \& {Aellig}}]{FaC01}
	{Fairfield}, D.~H., {Cairns}, I.~H., {Desch}, M.~D., {et~al.} 2001, \jgr, 106,
	25361
	
	\bibitem[{{Farris} {et~al.}(1991){Farris}, {Petrinec}, \& {Russell}}]{FPR91}
	{Farris}, M.~H., {Petrinec}, S.~M., \& {Russell}, C.~T. 1991, \grl, 18, 1821
	
	\bibitem[{{Formisano}(1979)}]{For79}
	{Formisano}, V. 1979, Nuovo Cimento C Geophysics Space Physics C, 2, 681
	
	\bibitem[{{Fujiki} {et~al.}(2016){Fujiki}, {Tokumaru}, {Hayashi}, {Satonaka},
		\& {Hakamada}}]{FuT16}
	{Fujiki}, K., {Tokumaru}, M., {Hayashi}, K., {Satonaka}, D., \& {Hakamada}, K.
	2016, \apjl, 827, L41
	
	\bibitem[{{Garc{\'{\i}}A} \& {Hughes}(2007)}]{GHu07}
	{Garc{\'{\i}}A}, K.~S., \& {Hughes}, W.~J. 2007, Journal of Geophysical
	Research (Space Physics), 112, A06229
	
	\bibitem[{{Gopalswamy} {et~al.}(2016){Gopalswamy}, {Yashiro}, \&
		{Akiyama}}]{GYa16}
	{Gopalswamy}, N., {Yashiro}, S., \& {Akiyama}, S. 2016, \apjl, 823, L15
	
	\bibitem[{{Janardhan} \& {Alurkar}(1993)}]{JAl93}
	{Janardhan}, P., \& {Alurkar}, S.~K. 1993, \aap, 269, 119
	
	\bibitem[{{Janardhan} {et~al.}(2011){Janardhan}, {Bisoi}, {Ananthakrishnan},
		{Tokumaru}, \& {Fujiki}}]{JaB11}
	{Janardhan}, P., {Bisoi}, S.~K., {Ananthakrishnan}, S., {Tokumaru}, M., \&
	{Fujiki}, K. 2011, \grl, 38, L20108
	
	\bibitem[{{Janardhan} {et~al.}(2015){Janardhan}, {Bisoi}, {Ananthakrishnan},
		{Tokumaru}, {Fujiki}, {Jose}, \& {Sridharan}}]{JaB15}
	{Janardhan}, P., {Bisoi}, S.~K., {Ananthakrishnan}, S., {et~al.} 2015, Journal
	of Geophysical Research (Space Physics), 120, 5306
	
	\bibitem[{{Janardhan} {et~al.}(2010){Janardhan}, {Bisoi}, \& {Gosain}}]{JBG10}
	{Janardhan}, P., {Bisoi}, S.~K., \& {Gosain}, S. 2010, \solphys, 267, 267
	
	\bibitem[{{Janardhan} {et~al.}(2005){Janardhan}, {Fujiki}, {Kojima},
		{Tokumaru}, \& {Hakamada}}]{JaF05}
	{Janardhan}, P., {Fujiki}, K., {Kojima}, M., {Tokumaru}, M., \& {Hakamada}, K.
	2005, \jgr, 110, 8101
	
	\bibitem[{{Janardhan} {et~al.}(2008{\natexlab{a}}){Janardhan}, {Fujiki},
		{Sawant}, {Kojima}, {Hakamada}, \& {Krishnan}}]{JaF08}
	{Janardhan}, P., {Fujiki}, K., {Sawant}, H.~S., {et~al.} 2008{\natexlab{a}},
	\jgr, 113, 3102
	
	\bibitem[{{Janardhan} {et~al.}(2008{\natexlab{b}}){Janardhan}, {Tripathi}, \&
		{Mason}}]{JDM08}
	{Janardhan}, P., {Tripathi}, D., \& {Mason}, H.~E. 2008{\natexlab{b}}, \aa,
	488, L1
	
	\bibitem[{{Jel{\'{\i}}nek} {et~al.}(2012){Jel{\'{\i}}nek}, {N{\v e}me{\v c}ek},
		\& {{\v S}afr{\'a}nkov{\'a}}}]{JNS12}
	{Jel{\'{\i}}nek}, K., {N{\v e}me{\v c}ek}, Z., \& {{\v S}afr{\'a}nkov{\'a}}, J.
	2012, Journal of Geophysical Research (Space Physics), 117, A05208
	
	\bibitem[{{Jian} {et~al.}(2011){Jian}, {Russell}, \& {Luhmann}}]{JRL11}
	{Jian}, L.~K., {Russell}, C.~T., \& {Luhmann}, J.~G. 2011, \solphys, 274, 321
	
	\bibitem[{{Lin} {et~al.}(2010){Lin}, {Zhang}, {Liu}, {Wang}, \& {Gong}}]{LiZ10}
	{Lin}, R.~L., {Zhang}, X.~X., {Liu}, S.~Q., {Wang}, Y.~L., \& {Gong}, J.~C.
	2010, Journal of Geophysical Research (Space Physics), 115, A04207
	
	\bibitem[{{Lopez} {et~al.}(2011){Lopez}, {Merkin}, \& {Lyon}}]{LML11}
	{Lopez}, R.~E., {Merkin}, V.~G., \& {Lyon}, J.~G. 2011, Annales Geophysicae,
	29, 1129
	
	\bibitem[{{Lu} {et~al.}(2011){Lu}, {Liu}, {Kabin}, {Zhao}, {Liu}, {Zhou}, \&
		{Xiao}}]{LuL11}
	{Lu}, J.~Y., {Liu}, Z.-Q., {Kabin}, K., {et~al.} 2011, Journal of Geophysical
	Research (Space Physics), 116, A09237
	
	\bibitem[{{McComas} {et~al.}(2013){McComas}, {Angold}, {Elliott}, {Livadiotis},
		{Schwadron}, {Skoug}, \& {Smith}}]{McA13}
	{McComas}, D.~J., {Angold}, N., {Elliott}, H.~A., {et~al.} 2013, \apj, 779, 2
	
	\bibitem[{{McComas} {et~al.}(2003){McComas}, {Elliott}, {Schwadron}, {Gosling},
		{Skoug}, \& {Goldstein}}]{McE03}
	{McComas}, D.~J., {Elliott}, H.~A., {Schwadron}, N.~A., {et~al.} 2003, \grl,
	30, 1517
	
	\bibitem[{{Mead} \& {Beard}(1964)}]{MBe64}
	{Mead}, G.~D., \& {Beard}, D.~B. 1964, \jgr, 69, 1169?1179
	
	\bibitem[{{Nemecek} \& {Safrankova}(1991)}]{NSa91}
	{Nemecek}, Z., \& {Safrankova}, J. 1991, Journal of Atmospheric and Terrestrial
	Physics, 53, 1049
	
	\bibitem[{{Olson}(1969)}]{Ols69}
	{Olson}, W.~P. 1969, \jgr, 74, 5642
	
	\bibitem[{{Peredo} {et~al.}(1995){Peredo}, {Slavin}, {Mazur}, \&
		{Curtis}}]{PeS95}
	{Peredo}, M., {Slavin}, J.~A., {Mazur}, E., \& {Curtis}, S.~A. 1995, \jgr, 100,
	7907
	
	\bibitem[{{Petrinec} \& {Russell}(1996)}]{PRu96}
	{Petrinec}, S.~M., \& {Russell}, C.~T. 1996, \jgr, 101, 137
	
	\bibitem[{{Readhead} \& {Hewish}(1972)}]{RHe72}
	{Readhead}, A.~C.~S., \& {Hewish}, A. 1972, \nat, 236, 440
	
	\bibitem[{{Richardson} {et~al.}(2001){Richardson}, {Wang}, \&
		{Paularena}}]{RWP01}
	{Richardson}, J.~D., {Wang}, C., \& {Paularena}, K.~I. 2001, Advances in Space
	Research, 27, 471
	
	\bibitem[{{Riley} {et~al.}(2015){Riley}, {Lionello}, {Linker}, {Cliver},
		{Balogh}, {Beer}, {Charbonneau}, {Crooker}, {DeRosa}, {Lockwood}, {Owens},
		{McCracken}, {Usoskin}, \& {Koutchmy}}]{RiL15}
	{Riley}, P., {Lionello}, R., {Linker}, J.~A., {et~al.} 2015, \apj, 802, 105
	
	\bibitem[{{Rout} {et~al.}(2017){Rout}, {Chakrabarty}, {Janardhan}, {Sekar},
		{Vrunda}, , \& {Pandey}}]{RoC17}
	{Rout}, D., {Chakrabarty}, D., {Janardhan}, P., {et~al.} 2017, \grl, L
	
	\bibitem[{{S{\'a}nchez-Sesma}(2016)}]{San16}
	{S{\'a}nchez-Sesma}, J. 2016, Earth System Dynamics, 7, 583
	
	\bibitem[{{Schatten} \& {Pesnell}(1993)}]{SPe93}
	{Schatten}, K.~H., \& {Pesnell}, W.~D. 1993, \grl, 20, 2275
	
	\bibitem[{{Shue} {et~al.}(1997){Shue}, {Chao}, {Fu}, {Russell}, {Song},
		{Khurana}, \& {Singer}}]{ShC97}
	{Shue}, J.-H., {Chao}, J.~K., {Fu}, H.~C., {et~al.} 1997, \jgr, 102, 9497
	
	\bibitem[{{Shue} \& {Song}(2002)}]{SSo02}
	{Shue}, J.-H., \& {Song}, P. 2002, \planss, 50, 549
	
	\bibitem[{{Shue} {et~al.}(1998){Shue}, {Song}, {Russell}, {Steinberg}, {Chao},
		{Zastenker}, {Vaisberg}, {Kokubun}, {Singer}, {Detman}, \& {Kawano}}]{ShS98}
	{Shue}, J.-H., {Song}, P., {Russell}, C.~T., {et~al.} 1998, \jgr, 103, 17691
	
	\bibitem[{Shukhtina \& Gordeev(2015)}]{SGo15}
	Shukhtina, M.~A., \& Gordeev, E. 2015, Annales Geophysicae, 33, 769
	
	\bibitem[{{Sibeck} {et~al.}(1991){Sibeck}, {Lopez}, \& {Roelof}}]{SLR91}
	{Sibeck}, D.~G., {Lopez}, R.~E., \& {Roelof}, E.~C. 1991, \jgr, 96, 5489
	
	\bibitem[{{Spreiter} \& {Briggs}(1962)}]{SBr62}
	{Spreiter}, J.~R., \& {Briggs}, B.~R. 1962, \jgr, 67, 37
	
	\bibitem[{{Spreiter} {et~al.}(1966){Spreiter}, {Summers}, \& {Alksne}}]{SSA66}
	{Spreiter}, J.~R., {Summers}, A.~L., \& {Alksne}, A.~Y. 1966, \planss, 14, 223
	
	\bibitem[{{T{\'o}th} {et~al.}(2005){T{\'o}th}, {Sokolov}, {Gombosi}, {Chesney},
		{Clauer}, {De Zeeuw}, {Hansen}, {Kane}, {Manchester}, {Oehmke}, {Powell},
		{Ridley}, {Roussev}, {Stout}, {Volberg}, {Wolf}, {Sazykin}, {Chan}, {Yu}, \&
		{K{\'o}ta}}]{ToS05}
	{T{\'o}th}, G., {Sokolov}, I.~V., {Gombosi}, T.~I., {et~al.} 2005, \jgr, 110,
	12226
	
	\bibitem[{{Usoskin} {et~al.}(2007){Usoskin}, {Solanki}, \& {Kovaltsov}}]{USK07}
	{Usoskin}, I.~G., {Solanki}, S.~K., \& {Kovaltsov}, G.~A. 2007, \aap, 471, 301
	
	\bibitem[{{Usoskin} {et~al.}(2014){Usoskin}, {Hulot}, {Gallet}, {Roth},
		{Licht}, {Joos}, {Kovaltsov}, {Th{\'e}bault}, \& {Khokhlov}}]{UsH14}
	{Usoskin}, I.~G., {Hulot}, G., {Gallet}, Y., {et~al.} 2014, \aap, 562, L10
	
	\bibitem[{{Verigin} {et~al.}(1999){Verigin}, {Kotova}, {Remizov}, {Styazhkin},
		{Schutte}, {Zhang}, {Riedler}, {Rosenbauer}, {Szego}, {Tatrallyay}, \&
		{Schwingenschuh}}]{VeK99}
	{Verigin}, M.~I., {Kotova}, G.~A., {Remizov}, A.~P., {et~al.} 1999, Cosmic
	Research, 37, 34
	
	\bibitem[{{Wang} {et~al.}(2013){Wang}, {Sibeck}, {Merka}, {Boardsen},
		{Karimabadi}, {Sipes}, {{\v S}afr{\'a}nkov{\'a}}, {Jel{\'{\i}}nek}, \&
		{Lin}}]{WaS13}
	{Wang}, Y., {Sibeck}, D.~G., {Merka}, J., {et~al.} 2013, Journal of Geophysical
	Research (Space Physics), 118, 2173
	
	\bibitem[{{Yang} {et~al.}(2003){Yang}, {Chao}, {Dmitriev}, {Lin}, \&
		{Ober}}]{YaC03}
	{Yang}, Y.-H., {Chao}, J.~K., {Dmitriev}, A.~V., {Lin}, C.-H., \& {Ober}, D.~M.
	2003, Journal of Geophysical Research (Space Physics), 108, 1104
	
	\bibitem[{{Zachilas} \& {Gkana}(2015)}]{ZGk15}
	{Zachilas}, L., \& {Gkana}, A. 2015, \solphys, 290, 1457
	
	\bibitem[{{Zhigulevsk} \& {Romishevskii}(1959)}]{ZRo59}
	{Zhigulevsk}, V.~N., \& {Romishevskii}, E.~A. 1959, Soviet Phys. Doklady, 5,
	1001
	
	\bibitem[{{Zhong} {et~al.}(2014){Zhong}, {Wan}, {Wei}, {Fu}, {Jiao}, {Rong},
		{Chai}, \& {Han}}]{ZhW14}
	{Zhong}, J., {Wan}, W.~X., {Wei}, Y., {et~al.} 2014, Journal of Geophysical
	Research (Space Physics), 119, 9816
	
	\bibitem[{{Zolotova} \& {Ponyavin}(2014)}]{ZPo14}
	{Zolotova}, N.~V., \& {Ponyavin}, D.~I. 2014, Journal of Geophysical Research
	(Space Physics), 119, 3281
	
\end{thebibliography}

\end{document}